\documentclass[aps,prl,reprint]{revtex4-1}

\usepackage{blindtext}
\usepackage{graphicx}
\usepackage[export]{adjustbox}
\usepackage{dcolumn}
\usepackage{bm}
\usepackage{amssymb}
\usepackage{amsmath}

\DeclareMathOperator{\diag}{diag}

\DeclareMathOperator{\Tr}{Tr}

\newcommand{\RN}[1]{%
\textup{\uppercase\expandafter{\romannumeral#1}}%
}

\usepackage{xcolor}

\begin{document}
\title{Multi-Relevance: Coexisting but Distinct Notions of Scale in Large Systems}

\author{Adam G. Kline}
\affiliation{%
 Department of Physics, The University of Chicago, Chicago IL 60637
}
\author{Stephanie E. Palmer}%
\affiliation{%
 Department of Organismal Biology and Anatomy and Department of Physics, The University of Chicago, Chicago IL 60637
}%

\begin{abstract}

Renormalization group (RG) methods are emerging as tools in biology and computer science to support the search for simplifying structure in distributions over high-dimensional spaces. We show that mixture models can be thought of as having multiple coexisting, exactly independent RG flows, each with its own notion of scale. We define this property as ``multi-relevance''. As an example, we construct a model that has two distinct notions of scale, each corresponding to the state of an unobserved categorical variable. In the regime where this latent variable can be inferred using a linear classifier, the vertex expansion approach in non-perturbative RG can be applied successfully but will give different answers depending the choice of expansion point in state space. In the regime where linear estimation of the latent state fails, we show that the vertex expansion predicts a decrease in the total number of relevant couplings from four to three and does not admit a good polynomial truncation scheme. This indicates oversimplification. One consequence of this is that principal component analysis (PCA) may be a poor choice of coarse-graining scheme in multi-relevant systems, since it imposes a notion of scale which is incorrect from the RG perspective. Taken together, our results indicate that RG and PCA can lead to oversimplification when multi-relevance is present and not accounted for.

\end{abstract}

\maketitle

Recently, renormalization group (RG) methods have been used in fields at the boundaries of traditional physics \cite{berman_inverse_2022, goldman_exact_2023, erbin_non-perturbative_2022, lahoche_generalized_2021, mehta_exact_2014, koch-janusz_mutual_2018, di_sante_deep_2022, balog_critical_2022, beny_renormalisation_2014, cotler_renormalization_2023, strandkvist_beyond_2020, pessoa_exact_2018, jentsch_critical_2023}. In the domain of theoretical biology, and in particular neuroscience, there have been attempts to apply RG methods to data \cite{bradde_pca_2017, meshulam_coarse_2019} and models \cite{brinkman_non-perturbative_2023, tiberi_gell-mannlow_2022} in order to find simplifying structure through the language of many-body physics. Yet many biological systems are more complex than field theories, and when directly adopting RG techniques there is danger of oversimplifying, or of failing to define important collective variables correctly. This raises the question of how one should choose a cutoff scheme when applying RG in biological systems.

To frame the problem, let us consider the firing patterns of $N$ neurons in a biological neural network. The joint distribution, $P(\sigma)$, is a probability distribution over $\sigma \in \{0,1\}^N$, and $-\log P(\sigma)$ is analogous to an energy for classical Ising spins. Models for this system have been fit to real data taken from the vertebrate retina, and the resulting energy landscapes have many local minima when constrained to a fixed total firing rate \cite{tkacik_searching_2014, prentice_error-robust_2016}. Because each of these collective states is thought to encode a different set of features in the stimulus, we expect the definitions of important collective variables to change depending on the region of state space under consideration. The problem is that most RG approaches implement linear coarse-graining, meaning a single collective basis is applied globally. Using the non-perturbative RG (NPRG), these statements can be made precise for distributions over finitely many variables.

In the Wetterich construction of NPRG, the cutoff is enforced by a adding a regularization term to the Hamiltonian which limits fluctuations in IR variables. For continuous degrees of freedom $\phi$, the Helmholtz energy is given by
\begin{equation}\label{eq:free_energy}
    W_k(J) = \log \int d\phi\, e^{-\mathcal{H}(\phi) - \frac{1}{2}\phi\cdot R_k \cdot \phi + J\cdot\phi},
\end{equation}
where, $R_k$ adds mass to long-wavelength collective variables ($|q|<k$), suppressing their fluctuation contributions to the integral, but leaves short-wavelength modes ($|q|>k$) unaffected. As $k\to 0$, it is required that $R_k\to 0$, so that $W_k$ becomes the exact connected generating function. In practice, the goal is to compute the effective average action
\begin{equation} \label{eq:eaa}
    \Gamma_k(\varphi) = \max_J\left\{\varphi\cdot J - W_k(J)\right\} - \frac{1}{2}R_k\cdot \varphi^2\,,
\end{equation}
which obeys the exact flow equation \cite{wetterich_exact_1993}
\begin{equation} \label{eq:Wetterich_flow}
    \partial_k \Gamma_k(\varphi) = \frac{1}{2} \Tr \left\{\partial_k R_k \cdot \left[R_k + \Gamma_k^{(2)}(\varphi)\right]^{-1} \right\}\,.
\end{equation}
Here, $\varphi = \langle\phi \rangle$ is the flowing expectation value, defined as the variational derivative of $W_k(J)$ with respect to $J$. As $k\to 0$, $\Gamma_k$ approaches the effective potential, denoted $\Gamma$, sometimes referred to as the Gibbs energy. NPRG methods are quite diverse in implementation and motivation (c.f.\ these reviews for broader background \cite{delamotte_introduction_2012,kopietz_introduction_2010,bagnuls_exact_2001,dupuis_nonperturbative_2021}).

The fact that this RG procedure implements linear coarse-graining can be seen in the quadratic dependence of the regulator term on $\phi$ in Eq.\ \eqref{eq:free_energy}. On the other hand, a coarse-graining scheme which captures biologically relevant features might use nonlinear coarse-graining \cite{kline_gaussian_2022}, but this would add $\phi$ dependence to the regulator $R_k$ and invalidate the Wetterich Eq.\ \eqref{eq:Wetterich_flow}. What is needed is some way to coarse-grain according to a number of different linear collective bases, without making the calculation intractable.

We provide a simple mechanism that allows for this kind of variability while leaving the essential flow equations unchanged. Some systems are most naturally modeled by a collection of coexisting, exactly independent RG flows, where each describes the collective physics according to a different notion of scale. We term systems with this property \textit{multi-relevant}.

A system described by a state $\phi$ and Hamiltonian $\mathcal{H}(\phi)$ is multi-relevant when there exists a finite set of Hamiltonians $\{\mathcal{H}_z(\phi)\}$ which are polynomial in $\phi$ and which have finite partition functions such that
\begin{equation}\label{eq:multi_relevance}
    \exp \left(-\mathcal{H}(\phi)\right) = \sum_z \exp\left( -\mathcal{H}_z(\phi)\right)\,.
\end{equation}
This expression, $\exp(-\mathcal{H})$, describes a mixture model whose components are normalizable and can be described by power-series expansions. Note that $z$ can be interpreted as a label on a latent state that is not explicitly part of the system state. An immediate consequence of this condition is that the total partition function is simply a sum of the component partition functions \footnote{A similar mixture construction has been used to perform accurate modeling of complex systems in which metastable states have a significant impact on the overall thermodynamics \cite{liu_zentropy_2022, liu_multiscale_2019}.}. This can be expressed in terms of the effective actions as $\exp \mathcal{L}[\Gamma](J) = \sum_{z} \exp \mathcal{L}[\Gamma_z](J)$, where $\mathcal{L}[F](J) = \max_\varphi\left\{\varphi \cdot J - F(\varphi)\right\}$ is the Legendre transform of $F$. By $\Gamma_z$ we mean the effective potential obtained from $\mathcal{H}_z$ by running the RG flow \eqref{eq:Wetterich_flow} down to $k\to0$. Because of this, each component model $\mathcal{H}_z$ can be renormalized independently subject to its own cutoff scheme $R_{z,k}$. The resulting effective potentials can be combined to give the full effective potential. In this sense, a multi-relevant system has multiple exactly independent RG flows.


We construct a toy model that is multi-relevant and has two mixture components, $\mathcal{H}_A$ and $\mathcal{H}_B$. Each is a finite, nonlocal analogue of scalar $\phi^4$ theory, centered about a point $s_z$, with $\phi\in\mathbb{R}^N$ and $N$ large. Explicitly,
\begin{align} \label{eq:phi_4_hamiltonian}
    \mathcal{H}_z(\phi+s_z) &= \sum_{ab}\frac{1}{2} (K_{ab}^z + u_{2}^z\delta_{ab})\phi_a\phi_b + \frac{u_{4}^z}{4! } \phi_a^2\phi_b^2\\
    \mathcal{H}(\phi) &= -\log\left[e^{-\mathcal{H}_A(\phi)} + e^{-\mathcal{H}_B(\phi)}\right]  \,. \label{eq:mixture_model}
\end{align}

We refer to \eqref{eq:mixture_model} as the ``$\phi^4$ mixture model''. Here, the usual kinetic term has been replaced by a generic positive semi-definite matrix $K_{ab}^z$ plus a ``mass'' part $u_{2}^z \delta_{ab}$, which is defined by requiring that the smallest eigenvalue of $K^z$ is unity. The eigenvector matrix $V^z$ of $K^z$ is sampled from the Haar measure on $O(N)$, and $s_z$ has a random orientation.

Although no notion of locality is present, we can still use RG, following \cite{bradde_pca_2017}. When analyzing the component energy labeled $z$, we take the collective degrees of freedom to be the eigenvectors of $K^z$. Assume that the $K^z$ eigenvalues are well-described by a density $\rho_z(\lambda)$. In the collective basis, the scale-free part of the kinetic term has the form:
\begin{equation*}
    \sum_{a}\lambda_{z,a} \phi_a^2 = \int d\lambda \rho_z(\lambda) \lambda \phi(\lambda)^2 \sim \int \frac{d^{D_z} q}{(2\pi)^{D_z/2}} q^2 |\phi(q)|^2
\end{equation*}
By assumption, $\rho_z$ depends on $\lambda$ as a power law, and we can identify the exponent with spatial dimensionality:
\begin{equation*}
    \rho_z(\lambda) \sim \lambda^{\alpha_z - 1} \Rightarrow \alpha_z = D_z/2
\end{equation*}

To be clear, we have not introduced anything like a spatial manifold in which these degrees of freedom are localized; this relation to dimension is only an analogy. The spectral density exponent $\alpha$ controls the scaling properties of couplings in the same way that dimensionality does in field theory. The UV scale $\Lambda_z$ is the largest eigenvalue of $K^z$ and scales as $N^{1/\alpha_z}$.

For our analysis, we use the vertex expansion method with expansion points $s_z$. Although the NPRG formalism allows for all entries of the effective vertices to be tracked, we perform a low-dimensional parameterization in line with the parameters of the original Hamiltonian. Our cutoff scheme is the Litim regulator \cite{litim_optimized_2001}, without field-strength renormalization. For component $z$, this this takes the form
\begin{equation}\label{eq:litim}
    (V_z^TR_{z,k}V_z)_{ab} = \max \left\{k - \lambda_{z,a}, 0 \right\}\delta_{ab}\,.
\end{equation}

The ansatz for the flowing effective action contains two couplings $u_{2,k}^z$ and $u_{4,k}^z$ for each component $z$:
\begin{align*}
    \Gamma^{(2)}_{z,k\,ab} &= K_{ab}^z + u_{2,k}^z \delta_{ab}\\
    \Gamma^{(4)}_{z,k\,abcd} &= \frac{u_{4,k}^z}{3}(\delta_{ab}\delta_{cd} + \delta_{ac}\delta_{bd} + \delta_{ad}\delta_{bc})
\end{align*}
The full RG analysis is given in Supplemental Material (SM) and is largely standard. For $\alpha_z \in (3/2,2)$, (correspondingly $D_z \in (3,4)$) the mass and interaction couplings $u_{z,2,k}$ and $u_{z,4,k}$ are relevant and describe the asymptotic properties. They obey the usual flow equations \cite{kopietz_introduction_2010} , up to differences in numerical factors due to neglecting sub-extensive diagrams. There are two fixed points, one at the origin and one at the WF fixed point. There is a $\mathbb{Z}_2$ symmetry-broken phase with two degenerate minima of the effective average action and a symmetric phase with a single minimum at $s_z$.

Because the $\phi^4$ mixture model is multi-relevant, $\Gamma$ can be obtained from the $k\to0$ limits of $\Gamma_{A,k}$ and $\Gamma_{B,k}$. These individual analyses of $A$ and $B$ are simple, including only two relevant parameters each. When combined to compute the whole effective potential at the end, there are still only four relevant parameters which constrain measurements (averages). Yet, the functional dependence of these measurements on the four parameters is complicated due to the mixture construction. By contrast, it is clear that if we had not used the mixture construction and instead computed the vertex expansion flow for the whole $\Gamma_k$, our final answer would look a lot simpler, namely it would be some polynomial in the fields. This simpler representation is wrong in some regimes, since it cannot capture the essentially non-polynomial dependence of $\Gamma(\phi)$ on $\phi$, due to the mixture construction. Despite this, there are regimes where the standard approach is justified.

\begin{figure}
    \centering
    \includegraphics[width=8.6cm]{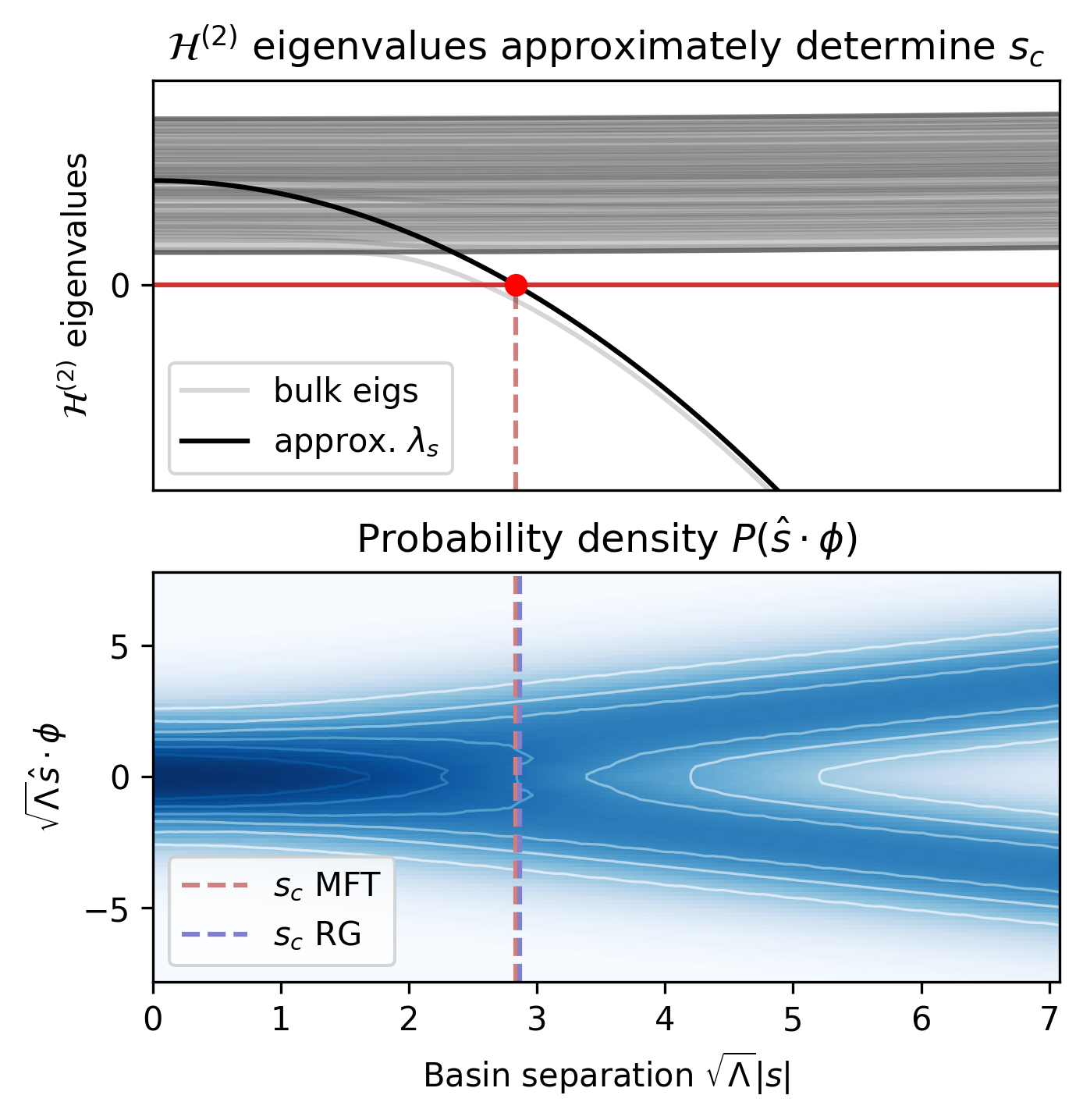}
    \caption{Demonstration of the coalescence transition. \textbf{Top}: spectrum of the energy Hessian at the saddle point separating basins $A$ and $B$. Model parameters are $N=50$, $\alpha_A=3/2$, $\Lambda_A=N^{1/\alpha_A}$, $u_2^A = -\Lambda_A/10$, $u_4^A = 0.58 \Lambda_A^{s_4}$. All $A$ and $B$ parameters are equal. The gray lines depict the true eigenvalues (neglecting $O(|s|^4)$ effects) while the black line represents $\lambda_s = \hat{s} \cdot \mathcal{H}^{(2)} \cdot \hat{s}$. The separation $|s|$ at which $\lambda_s$ crosses 0 is the mean-field estimate. \textbf{Bottom}: probability density of the $\phi\cdot\hat{s}$ as a function of separation. Picking a value on the $x$-axis selects a single distribution with a specified separation. The red dashed line indicates the mean-field estimate of the critical separation, while the blue dashed line gives the RG estimate (see SM for expressions).}
    \label{fig:coalescence_transition}
\end{figure}

Whether standard vertex expansion succeeds or fails is essentially determined by the parameters $s_A$ and $s_B$. These act as displacements in state space, centering one $\phi^4$ model at $s_A$ and the other at $s_B$. Let $s = s_A - s_B$ be the separation vector. In Fig.\ \ref{fig:coalescence_transition}, we show the separation dependence of the probability density of $\phi$ projected along the inter-basin axis $\hat{s}$. There is a transition as $|s|$ crosses some critical threshold $s_c$, which we calculate in the SM. To be concrete, we define this transition as the largest separation at which $P(\hat{s} \cdot \phi)$ has a local maximum at $\hat{s} \cdot \phi = 0$. This approximates the separation beyond which the $A$ and $B$ component densities can be separated by a hyperplane.

Consider first the well-separated phase, that is $|s|>s_c$. The total energy has a saddle point near $s_A$, and largely the local structure looks like $\mathcal{H}_A$. This is because the coefficient data due to the presence of $\mathcal{H}_B$ are suppressed exponentially with respect to a power of the separation $|s|$. For states $\phi$ near $s_A$,
\begin{align*}
    \mathcal{H}(\phi) &= \mathcal{H}_{A}(\phi) - e^{- \Delta \mathcal{H}_{BA}(\phi)} + O\left(e^{-2\Delta \mathcal{H}_{BA}(\phi)}\right),
\end{align*}
where $\Delta \mathcal{H}_{AB} = \mathcal{H}_B - \mathcal{H}_A$. Then, using the regulator $R_{A,k}$ in \eqref{eq:litim}, one can compute $\Gamma_k$ in the neighborhood of $s_A$, using the above as the initial conditions of the flow. For sufficiently large separations $|s| > s_c$, the perturbations to $\mathcal{H}_A$ due to $\mathcal{H}_B$ do not change the fact that the calculation of $\Gamma_k$ recovers a good approximation of $\Gamma_{A,k}$. This follows from the fact that when the basins are well-separated, a series expansion of $\Gamma(\phi)$ will approximate that of $\Gamma_z(\phi)$ for $\phi$ in a suitable neighborhood of $s_z$. In turn, this occurs because the Helmholtz free energy for one component, $W_z(J)$, dominates the mixture family for $J$ near some $J^*$, making $\Gamma(\phi)\approx\Gamma_z(\phi)$ for $\phi$ in the neighborhood of $\phi(J^*)$. We discuss this further in the SM.

The vertex expansion approach can therefore, in the well-separated case, be sufficient to analyze a multi-relevant energy function $\mathcal{H}$. However, to get a full description of $\Gamma$, one needs to perform several RG flows, one at each $s_z$. In general, each requires a different regulator scheme, and will give different asymptotic properties than the others. In practice, this is equivalent to using the mixture construction. It is interesting to note that this analysis demonstrates how multi-relevant models can have differing low-energy physics near different points in state space. If there were, additionally, some notion of dynamics that led to effective ergodicity breaking, one could consider the system getting trapped in basin $A$ or $B$ for long periods of time, and in each case the other basin should have little influence.

In the coalesced phase, the lowest-order approximation to the vertex expansion approach fails. In this case, the separation $|s|$ is smaller than $s_c$. For simplicity, we consider the case $|s| = 0$. Expanding \eqref{eq:mixture_model} about the point $s_A = s_B = 0$ yields, at the quadratic level,
\begin{equation*}
    \mathcal{H}^{(2)}_{ab} = K_{ab} + \frac{1}{2}(u_{2}^A + u_{2}^B)\delta_{ab}\,,
\end{equation*}
where $K = (K^A + K^B)/2$. Because $K^A$ and $K^B$ are diagonal in different, randomly chosen bases, their sum $2K$ defines a new notion of scale that does not completely agree with either system $A$ or $B$. As before, we define scale using the eigenvalues of $K$, with the collective degrees of freedom (analogous to Fourier modes) defined by the eigenvectors. In our construction, $K$ acquires a mass gap $K_0$, and after subtracting this off, the remaining eigenvalues will have some asymptotically scale-free behavior. Therefore let $\{\lambda_a + K_0\}_{a=1}^N$ denote the eigenvalues of $K$, and construct the density $\rho(\lambda)$ of these eigenvalues as we did in the finite $\phi^4$ case. The units associated to these eigenvalues are defined through the scaling properties of this density. For example, since $\rho(\lambda) \sim \lambda^{\alpha - 1}$, an extensive quantity has dimension $\alpha$.

Expanding \eqref{eq:mixture_model} to higher orders in $\phi$, all of the terms come in the form of symmetrized outer products involving $K^A$, $K^B$, and $\delta$, with $\delta$ the identity matrix. For example, the $\mathcal{H}^{(4)}$ has a term like $(K^{A})_{(ab} (K^{B})_{cd)}$, which we denote $(K^A \circ K^B)_{abcd}$. In the eigenbasis of $K$, the matrices $K^A$ and $K^B$ can be approximated as diagonal, but the diagonal elements pick up a mass gap just as $K$ itself does. By defining $\tilde{K}^A = K^A - K^A_0 \delta$ as the scale-free part of the $K^A$ diagonal elements (and repeating for $K^B$), we can write the expansion in terms of approximately scale-free objects. After computing scaling dimensions, the relevant and marginal couplings are captured by the ansatz:
\begin{align}
    \Gamma_{k\,ab}^{(2)} &= (v^A_k \tilde{K}^A + v^B_k \tilde{K}^B + v^\delta_k \delta)_{ab} \label{eq:quadratic_ansatz} \\
    \Gamma_{k\,abcd}^{(4)} &=  v^{\delta\delta}_k (\delta\circ\delta)_{abcd} \label{eq:quartic_ansatz}\\
    \Gamma^{(6)}_{k\,a\dots f} &= v^{\delta \delta \delta}_k (\delta \circ \delta \circ \delta)_{a\dots f} \label{eq:sixtic_ansatz}
\end{align}
The spectral density exponent $\alpha$ determined by this new quadratic coupling $K$ is smaller than $\alpha_A$ and $\alpha_B$ when we take them to be equal. Because of this, the sixtic interaction becomes relevant and must be included. Yet, for $\alpha_A = \alpha_B$ in the range $(3/2,2)$, no other terms appearing in the initial conditions generated by the expansion of $\mathcal{H}$ have positive scaling dimensions at the Gaussian fixed point. The couplings we include provide three unique constraints on the original model parameters, and the marginal coupling initial conditions provide no additional constraints.

To compute the flow equations, we employ the Litim regulator (without field-strength renormalization) as before. Keeping only extensive terms yields
\begin{align*}
    \dot{g}^\delta_l &= g^{\delta}_l + \frac{1}{6} \frac{g^{\delta\delta}_l}{(1+g^{\delta}_l)^2}\\
    \dot{g}^{\delta\delta}_l &= (2-\alpha)g^{\delta\delta}_l - \frac{1}{3} \frac{(g^{\delta\delta}_l)^2}{(1+g^{\delta}_l)^3} + \frac{1}{10}\frac{g^{\delta\delta\delta}}{(1 + g^\delta)^2}\\
    \dot{g}^{\delta\delta\delta}_l &= (3-2\alpha)g^{\delta\delta\delta}_l + \frac{5}{3} \frac{(g_l^{\delta\delta})^3}{(1+g_l^\delta)^4} - \frac{g_l^{\delta\delta}\,g_l^{\delta\delta\delta}}{(1+g_l^\delta)^3}
\end{align*}
where $l = -\log(k/\Lambda)$ and $g$ are the non-dimensionalized couplings. These flow equations yield three finite fixed points: The Gaussian point at the origin, the WF point, and one other point that appears for $\alpha\leq 3/2$. In effect, a standard application of the vertex expansion approach predicts that the IR properties of the mixture model in the coalesced phase are the same as some kind of nonlocal $\phi^6$ model near and below 3 dimensions (but still above $D = 8/3$ where $\phi^8$ becomes relevant).

The crucial result that we come to now is that the vertex expansion approach on the full model $\mathcal{H}$ at $s_A=s_B$ led to oversimplification. A primary indication of this is there are only three relevant parameters, while according to the separate or mixture analysis there should be four. In systems with increasing $N$, the net effect of irrelevant couplings on the large-scale components of average quantities decreases, and the possibility of constraining their initial values using data disappears. By contrast, we know that all four parameters, $u_{z,2}$ and $u_{z,4}$, can affect predictions if the mixture construction is used, regardless of the basin separation $|s|$. This disagreement arises at the choice of truncation scheme \eqref{eq:quadratic_ansatz}, \eqref{eq:quartic_ansatz}, and \eqref{eq:sixtic_ansatz}. Due to the non-polynomial structure of the microscopic model $\mathcal{H}$, an infinite number of irrelevant couplings with large initial values are thrown away. However, these have a significant effect on the flow of relevant and marginal couplings in the UV and, for accurate predictions, must be included. In essence, the single-flow approach oversimplifies because its notion of scale is poorly chosen and no good polynomial truncation scheme is available.



An important consequence of this calculation is that when principal component analysis (PCA) is applied to data drawn from a multi-relevant distribution, it can also lead to oversimplification. In data from our model, one finds approximately power-law distributed covariance eigenvalues and increasingly non-Gaussian behavior in the IR, suggesting a non-trivial RG flow.
However, by performing a single RG flow starting at $s_A = s_B$ (or at the saddle point between the basins), we chose a collective basis and cutoff scheme which emulates PCA \cite{bradde_pca_2017, lahoche_generalized_2021}. Under coarse-graining with respect to this notion of scale, the RG analysis fails to effectively constrain the statistics of large-scale degrees of freedom using the original parameters. In contrast, the non-Gaussian statistics observed for large principal components is accounted for by the mixture model construction. Like the single-expansion point RG analysis, PCA can lead to oversimplification because it mixes up the collective bases defined by $K^A$ and $K^B$, and after coarse-graining their effects cannot be disentangled.

In this letter, we introduced a way to formalize multi-relevance. This property endows a system with multiple coexisting and exactly independent RG flows, each with its own potentially distinct notion of scale. The existence of latent categorical variables is a simple mechanism which leads directly to multi-relevance, essentially by definition. We expect these results to be an important step towards implementing RG techniques in theoretical biology and complex systems more broadly. In these new domains, RG could offer paths to simplification in the space of models and thereby aid in the search for organizing principles. Our analysis reveals that multi-relevance does not simplify under the RG flow, and in this sense represents a degree of inherent complexity which must be accounted for when building and classifying theories. Whereas PCA fails to correctly coarse-grain multi-relevant systems, machine learning methods have demonstrated promise in discovering latent representations together with rich, nonlinear encoding and decoding schemes \cite{hinton_reducing_2006, berman_predictability_2016, ding_deciphering_2019, ahamed_capturing_2021, kalinin_exploring_2021, ziegler_latent_2023}. Through multi-relevance, these capabilities are brought closer to many-body formalism and new theories of collective computation.




\section{Acknowledgements}

We would like to thank Cheyne Weis and Patrick Jentsch for useful feedback and discussions during the preparation of this manuscript. This work was partially supported by the University of Chicago Materials Research Science and Engineering Center, which is funded by the National Science Foundation under award number DMR-2011854. Aditionally, this work was supported by the National Science Foundation through the Center for the Physics of Biological Function (PHY-1734030), as well as the National Institutes of Health BRAIN initiative (R01EB026943).

\bibliography{references}

\appendix
\begin{widetext}
\section{Supplemental Material}

\subsection{NPRG on the finite $\phi^4$ model}\label{appendix:NPRG_on_finite_system}

In this section, we work through the NPRG analysis of a finite $\phi^4$ model as defined in the main text, Equation \ref{eq:phi_4_hamiltonian}:
\begin{equation*}
    \mathcal{H}(\phi) = \frac{1}{2} \sum_{ab}(K_{ab} + u_2 \delta_{ab}) \phi_a\phi_b + \frac{u_4}{4!}\phi_a^2 \phi_b^2
\end{equation*}

In Wetterich's formulation of the NPRG, the flow is defined in terms a slightly modified version of the 1PI generating functional, or Gibbs energy, known as the effective average action:
\begin{equation}\label{eq:effective_average_action}
    \Gamma_k(\varphi) = \max_J\left\{\varphi \cdot J - \log \int d\phi \exp\left[ -\mathcal{H}(\phi) - \frac{1}{2}R_{k\,ab}\,\phi_a\phi_b + J \cdot \phi\right]\right\} - \frac{1}{2} R_{k\,ab}\,\varphi_a\varphi_b
\end{equation}

The matrix $R$ is known as the regulator. It ensures that the the largest eigenvalue of the propagator is no larger than $1/k$ (or $1/k^2$, traditionally, though we choose an alternative parameterization for convenience). The full propagator at scale $k$ can be found using the effective action and the regulator:
\begin{equation*}
    \left[G^{-1}_{k}\right]_{ab} = \frac{\partial}{\partial \varphi_a} \frac{\partial}{\partial \varphi_b} \Gamma_k(\varphi) + R_{k\,ab}
\end{equation*}

It is important to note that unlike a hard cutoff scheme, the whole configurational integral over states $\phi$ in \eqref{eq:effective_average_action} is carried out, including fluctuations in IR modes. The regulator takes the role as the cutoff, and given that it satisfies the necessary boundary conditions, the resulting RG flow is well-defined for a finite number of degrees of freedom. This is not the case with a hard cutoff scheme, wherein one needs to rescale the total number of degrees of freedom in order to map the parameters of the coarse-grained Hamiltonian back in to the original space of couplings. This does not mean that there is not a rescaling step eventually, rather this new rescaling step happens after the flow has already been defined.

We use the vertex expansion, meaning the flow equations for vertices $\Gamma^{(n)}_{a_1,\dots,a_n}$ are found by differentiating the Wetterich flow equation directly and keeping track of expansion coefficients:
\begin{align*}
    \dot{\Gamma}_k (\varphi) &= \frac{1}{2} \Tr \left[ \dot{R}_k G_k(\varphi)  \right] \\
    \dot{\Gamma}_{ab}^{(2)} &= -\frac{1}{2} \Tr \left[\Gamma^{(4)}_{ab} G_k \dot{R}_k G_k\right] + \Tr \left[ \Gamma^{(3)}_a G_k \Gamma^{(3)}_b G_k \dot{R}_k G_k \right]\\
    \dot{\Gamma}_{abc}^{(3)} &= \dots \\
    \dot{\Gamma}_{abcd}^{(4)} &= -\frac{1}{2} \Tr \left[\Gamma^{(6)}_{abcd} G_k \dot{R}_k G_k\right] + 3 \Tr \left[ \Gamma^{(4)}_{ab} G_k \Gamma^{(4)}_{cd} G_k \dot{R}_k G_k \right]_{\text{(symmetrized)}} \\
    &\qquad + \text{terms involving } \Gamma^{(3)}
\end{align*}
From here on, we implicitly symmetrize where necessary. To evaluate these flow equations we need to truncate the hierarchy at some highest power $n$ and parameterize the vertices with an ansatz. It is easiest to start with a good parameterization and verify that higher-order vertices will not be relevant. Working from the high-temperature phase, all odd vertices such as $\Gamma^{(3)}$ are set to zero. The initial condition on $\Gamma_k$ is $\Gamma_{k\to\infty} = \mathcal{H}$. We approximate this by setting $\Gamma_\Lambda = \mathcal{H}$. Truncating at order $n=4$, the effective action can be written in the same form as $\mathcal{H}$:
\begin{align*}
    \Gamma^{(2)}_{k\,ab} &= (\lambda_a + u_{2\,k}) \delta_{ab} \\
    \Gamma^{(4)}_{k\,abcd} &= \frac{u_{4\,k}}{3}\left(\delta_{ab}\delta_{cd}+ \delta_{ac}\delta_{bd} + \delta_{ad}\delta_{bc}\right) \\
    u_{2\,\Lambda}&=u_2\\
    u_{4\, \Lambda}&=u_4
\end{align*}
From the truncation and $\phi\to-\phi$ symmetry we therefore have the following closed set of finitely many flow equations:
\begin{align*}
    \dot{\Gamma}_{ab}^{(2)} &= -\frac{1}{2} \Tr \left[\Gamma^{(4)}_{ab} G_k \dot{R}_k G_k\right]\\
    \dot{\Gamma}_{abcd}^{(4)} &= 3 \Tr \left[ \Gamma^{(4)}_{ab} G_k \Gamma^{(4)}_{cd} G_k \dot{R}_k G_k \right]
\end{align*}
To obtain a low-dimensional description of the flow, we re-express these flow equations in terms of the two parameters $u_{2\,k}$ and $u_{4\,k}$ of our ansatz. For example, we first insert $\Gamma^{(4)}$ into the RHS of the expression for $\dot{\Gamma}^{(2)}$  above.
\begin{align*}
     \partial_k \Gamma^{(2)}_{k\,ab} &= -\frac{1}{2} \frac{u_{4\,k}}{3} \sum_{cd}{(\delta_{ab}\delta_{cd}+ \delta_{ac}\delta_{bd} + \delta_{ad}\delta_{bc})\left(G_k \dot{R}_k G_k\right)_{cd}}\\
     &= -\frac{u_{4\,k}}{6} \left(\Tr\left(G_k \dot{R}_k G_k\right)\delta_{ab} + 2 \left(G_k \dot{R}_k G_k\right)_{ab} \right)\,.
\end{align*}
The first and second terms on the second line differ by a factor proportional to the number of degrees of freedom in the system, as we will show shortly. For this reason, we refer to the first term as extensive, while the second is sub-extensive. To actually calculate these terms, we need to introduce a regulator and get an expression for the propagator. In this context, the Litim regulator without field-strength renormalization can be written in the $\Gamma^{(2)}$ eigenbasis as
\begin{equation*}
    R_{k\,ab} = \max \left\{k - \lambda_a, 0 \right\}\delta_{ab}
\end{equation*}
This gives us
\begin{align*}
    \dot{R}_{k\,ab} &= 0\,;\quad a>n_k \\
    &=\delta_{ab}\,;\quad a<n_k \\
    G_{k\,ab} &= \left(\Gamma^{(2)}_k\right)^{-1}_{ab}\,; \quad a>n_k \\
    &=\frac{1}{k + u_{2\,k}}\delta_{ab}\,;\quad a>n_k
\end{align*}
and so
\begin{align*}
     \dot{\Gamma}^{(2)}_{k\,ab} &= -\frac{1}{2}\frac{u_{4\,k}}{3} \left(\frac{n_k}{(k + u_{2\,k})^2}\delta_{ab} +  \frac{2}{(k + u_{2\,k})^2}\dot{R}_{k\,ab} \right)
\end{align*}
By $n_k$ we mean the number of eigenvalues $\lambda_a$ below the cutoff $k$. We will mainly be interested in the intermediate part of the flow, for which $n_k$ is a number much larger than unity, even though it may be much less than $N$. This is the regime in which the scale-free parts of these flow equations are accurate and look similar to traditional results in infinite systems. Dropping the sub-extensive term and inserting our ansatz for $\Gamma^{(2)}$ into the RHS of the above equation, we find:
\begin{equation*}
     \dot{u}_{2\,k} = -\frac{1}{6}\frac{n_k u_{4\,k}}{(k + u_{2\,k})^2}
\end{equation*}
The computation for $\partial_k u_{4\,k}$ proceeds similarly. Inserting the effective action ansatz into the RHS of the truncated flow equations yields
\begin{align*}
    \dot{\Gamma}^{(4)}_{k\,abcd} &= 3 \Tr \left[ \Gamma^{(4)}_{ab} G_k \Gamma^{(4)}_{cd} G_k \dot{R}_k G_k \right] \\
    &= 3 u_{4\,k}^2 \sum_{a'b'c'd'} \frac{1}{3}(\delta_{ab}\delta_{a'b'} + \delta_{aa'}\delta_{bb'} + \delta_{ab'}\delta_{ba'}) \times\\
    &\qquad\qquad\qquad\:\: \frac{1}{3}(\delta_{cd}\delta_{c'd'} + \delta_{cc'}\delta_{dd'} + \delta_{cb'}\delta_{dc'})(G_k\dot{R}_{k}G_k)_{a'c'}\, G_{k\,b'd'}\\
    &= \frac{1}{3} u_{4\,k}^2\left( \delta_{ab}\delta_{cd} \Tr\left( G^3_k \dot{R}_k\right) + \text{sub-extensive terms}\right)\\
    &\to \frac{1}{3}\frac{n_k u_{4\,k}^2}{(k + u_{2\,k})^3} \delta_{ab}\delta_{cd}
\end{align*}
where the RHS is symmetrized across $abcd$. Substituting on the LHS, we have the scale-free part
\begin{equation*}
    \dot{u}_{4\,k} = \frac{1}{3}\frac{n_k}{N}\frac{u_{4\,k}^2}{(k + u_{2\,k})^3}
\end{equation*}
In this case, we can assign dimensions to the couplings simply by inspecting the flow equations. The key is to note that, for $[k] = 1$ by definition,
\begin{equation*}
    \alpha = \partial_{\log \lambda}\log\rho(\lambda) + 1 \quad \Rightarrow \quad [n_k] = \alpha\,.
\end{equation*}
A heuristic way to see this is to give $\lambda_n$ a simple expression consistent with $\rho(\lambda)\sim \lambda^{\alpha-1}$. This is satisfied by
\begin{equation*}
    \lambda_n \sim \Lambda\left(\frac{n}{N}\right)^{1/\alpha}\,.
\end{equation*}
Proceeding, we can straightforwardly identify:
\begin{equation*}
    s_2 = [u_2] = 1\quad s_4 = [u_4] = 2-\alpha
\end{equation*}
Finally, we define rescaled dimensionless couplings $g_2 = k^{-s_2}u_2$ and $g_4 = k^{-s_4}u_4$, and let $l = -\log(k/\Lambda)$. The dimensionless flow equations are given by:
\begin{align*}
    \partial_l g_{2\,l} &= g_{2\,l} + \frac{1}{6} \frac{g_{4\,l}}{(1+g_{2\,l})^2} \\
    \partial_l g_{4\,l} &= (2-\alpha) g_{4\,l} - \frac{1}{3} \frac{g_{4\,l}^2}{(1+g_{2\,l})^3}
\end{align*}
These have the standard properties describing the $\phi^4$ system, up to some differences in coefficients, and with anomalous scaling neglected. Save for the infinite fixed points, when $\alpha>2$ the only physical ($g_4\geq0$) fixed point is at $(0,0)$. For $\alpha<2$, the Wilson-Fisher fixed point appears at
\begin{equation*}
    g_2^* = \frac{2-\alpha}{4-\alpha}\qquad g_4^* = \frac{24(2-\alpha)}{(4-\alpha)^2}\,.
\end{equation*}

\subsection{Approximate behavior of $\Gamma$ in well-separated phase}

In the main text, we discussed that for a multi-relevant Hamiltonian $\mathcal{H}$ with component Hamiltonians $\mathcal{H}_z$, the full effective potential $\Gamma$ satisfies the exact relation
\begin{equation*}\label{eq:mixture_construction}
    \exp \mathcal{L}[\Gamma](J) = \exp W(J) = \sum_{z} \exp W_z(J) = \sum_{z} \exp \mathcal{L}[\Gamma_z](J)
\end{equation*}
In this section, we argue that in the well-separated phase, $\Gamma$ behaves like $\Gamma_z$ for $\phi$ in the neighborhood of $s_z$. Further, we argue that although $\Gamma$ is convex in principle, it may still be useful to consider it as the convex hull of a non-convex combination of the potentials $\Gamma_z$.

To begin, we work out a toy model. Consider a state variable $\phi \in \mathbb{R}$ which is distributed according to a mixture of Gaussians with unit variance and with means at $\pm a$. That is,
\begin{equation*}
    \mathcal{H}(x) =-\log \left[\exp(-\mathcal{H}_{+}(\phi)) + \exp(-\mathcal{H}_{-}(\phi))\right] = - \log \left[\exp\left(-\frac{1}{2}(\phi-a)^2\right) + \exp\left(-\frac{1}{2}(\phi+a)^2\right)\right]
\end{equation*}
The Helmholtz energy of $\mathcal{H}_{\pm}(\phi)$ is
\begin{equation*}
    W_{\pm}(J) = \log \int d\phi \exp(-\mathcal{H}_{\pm}(\phi) + J\phi) = \frac{1}{2} J^2 \pm J a
\end{equation*}
By applying a Legendre transform to these functions we get the effective potentials
\begin{equation*}
    \Gamma_z(\phi) = \frac{1}{2} (\phi - s_z)^2 = \mathcal{H}_z(\phi)
\end{equation*}
where $s_{\pm} = \pm a$. Define $\varphi = \partial_J W(J)$, the average of $\phi$ given source $J$. Then
\begin{equation*}
    \varphi = J + a \tanh Ja
\end{equation*}
So for $J>1/a$, $\varphi \to J + a$. As the separation $a$ is made larger, $1/a$ decreases, and this asymptotic behavior occurs for smaller values of $J$. To get a better intuition for this result, consider the full effective potential in this asymptotic case, where $a$ is large (relative to unity, the standard deviation of a single mixture component). When $J > 1/a$, the sum over $z$ is dominated by the term $\exp W_{+}(J)$. This yields:
\begin{equation*}
    \Gamma(\varphi) = \mathcal{L}[W](\varphi) \approx \mathcal{L}[W_+](\varphi) = \Gamma_+(\varphi)
\end{equation*}
In words, the full Helmholtz energy $W(J)$ is well-approximated by $W_{+}(J)$ for large positive sources $J$, i.e. $J > 1/a$ (and $W_{-}(J)$ for $J < -1/a$). Therefore, when we calculate $\Gamma$ from $W$ via Legendre transform, the fact that $W$ is dominated individual components $W_z$ in various regions causes $\Gamma$ to approximate $\Gamma_z$ in those regions.

How does this relate to our model and the definition of the critical separation $s_c$? First note that the toy model can be extended to a larger state space and non-Gaussian component energy functions. What is necessary is simply that the component densities are well-separated, so that $\langle\phi\rangle_J$ can take on large values at relatively small sources $J$. Next, our choice to define $s_c$ in terms of the maximum of the probability density of $\phi \cdot \hat{s}$ was not crucial. The important point is that the separation between the basins is greater than the variance (in the inter-basin direction) of each basin individually. In fact, we compute $s_c$ according to this condition later in the SM. At $J=0$, the second derivative of $W(J)$ with respect to $J$ gives the total covariance $\langle\phi^2\rangle - \langle \phi \rangle^2$. For large separations, this covariance value drops off quickly as $J$ is moved away from zero, since a slight source ``pushes'' the system into one basin or another. By assumption, these basins individually have much less variance than the total distribution at $J=0$.

Typically, this structure is associated with symmetry breaking and ergodicity breaking, wherein a small external source causes the system to find a state in an ergodically broken region. In this paper, we have only been concerned with distributions over finitely many variables, and adding dynamics would require additional assumptions and structure. However, given the close analogies to standard analysis in many-body systems, we believe that an effective description of the structure of $\Gamma(\phi)$ in the well-separated phase is provided by $\Gamma_z(\phi)$ when $\phi$ is near $s_z$, i.e., the minima of $\Gamma_z(\phi)$ can sometimes be treated as metastable states. While this manifestly breaks convexity of $\Gamma$, the convex hull can always be taken if needed, and otherwise $\Gamma$ can be interpreted as a constrained free energy.

\subsection{RG flow in the coalesced phase}

\subsubsection{Units and scaling dimensions}

In the previous section, the units of couplings $u_2$ and $u_4$ were not determined until after the flow equations were found. In general, this is the correct approach, namely that the units of different couplings should be verified by showing that they remove explicit scale-dependence in the flow equations. In the following discussion, the full (before irrelevant terms are removed) flow equation hierarchy is very complicated and a similar, `by inspection' approach is difficult. In this section, we discuss heuristic reasoning that provides scaling dimensions in terms of the original energy function. However, we wish to emphasize that the true scaling dimensions may not always appear when using this heuristic, since they are really defined by a cutoff-scheme and are revealed in the flow equations.

The basic strategy follows \cite{bradde_pca_2017} and \cite{lahoche_generalized_2021}, though we do not take into account non-power law dependence of $\rho$ on $\lambda$. As the cutoff scale $k$ is changed, the number $n_k$ of remaining modes below this cutoff scales like $n_k \sim k^\alpha$, which we denote by $[n_k] = \alpha$. What we would like to compute are the so-called `engineering dimensions' of our couplings, which can be defined as the critical exponents at the non-interacting (Gaussian) fixed point. First, consider the trace of the propagator under a change of $k$, where $k$ is the UV cutoff:
\begin{equation*}
    \Tr_k [G] = \sum_{a < n_k} \langle \phi_a^2 \rangle = \int_0^k d\lambda \,\rho(\lambda) \langle \phi(\lambda)^2 \rangle = \int_0^k d\lambda\,\rho(\lambda) \lambda^{-1} = \frac{\alpha N}{\Lambda^\alpha}\int_0^k d\lambda\, \lambda^{\alpha-2} \sim k^{\alpha - 1}
\end{equation*}
We conclude that
\begin{equation*}
    [\phi] = \frac{1}{2}(\alpha-1)
\end{equation*}
Note that under the identification $\lambda \sim q^2$ and $\alpha \sim D/2$ as pointed out in the Main Text, this agrees with the standard result $[\phi] \sim D/2 - 2$. Now consider a positive-definite matrix $C$ which is diagonal in the same basis as $\langle\phi_a \phi_b\rangle$, and with eigenvalues $\tilde{\lambda}_{a}$, where $a$ is ordered so that $\lambda_a$ are monotonically increasing. Let the density of $\tilde{\lambda}_a$ be denoted $\tilde{\rho}$, which integrates to $N$. Then define $\tilde{\alpha}(k)$ as
\begin{equation*}
    \tilde{\alpha} = \partial_{\log k} \log \tilde{n}_k\,;  \quad \tilde{n}_k = \int_0^k d\lambda\, \tilde{\rho}(\lambda)
\end{equation*}
This gives a mapping between the eigenvalues $\tilde{\lambda}_a$ of $C$ and the eigenvalues $\lambda_a$ of $K$ through $n_{\lambda} = \tilde{n}_{\tilde{\lambda}(\lambda)}$. When $\tilde{\rho}$ has power-law dependence on $\tilde{\lambda}$, $\tilde{n}_k \sim k^{\tilde{\alpha}}$, which yields:
\begin{equation*}
    \tilde{\lambda}(\lambda) \sim \lambda^{\alpha/\tilde{\alpha}} \equiv \lambda^\beta
\end{equation*}
Now the scaling dimension of $C$ can be computed in the same way:
\begin{equation*}
    \Tr_k[C G] = \sum_{a < n_k} \langle C_{aa}\phi_a^2 \rangle = \sum_a \lambda_a^{-1} \tilde{\lambda}_a  \sim \int_{0}^k d\lambda\, \rho(\lambda) \lambda^{\beta - 1}\sim k^{\alpha + \beta - 1}
\end{equation*}
Hence
\begin{equation*}
[C] = \frac{\alpha}{\tilde{\alpha}} = \beta
\end{equation*}
In our toy model, we use this to calculate the effective scaling dimensions of $\tilde{K}^A$ and $\tilde{K}^B$ in the mixed basis. We find that $\beta\approx 1$, meaning that, as far as power-counting rules are concerned, these matrices can be treated like derivative operators. Terms with higher powers of $\tilde{K}^{z}$ are increasingly irrelevant near the Gaussian fixed point.

Finally, let us consider a quartic coupling comprised of the symmetrized outer product of two matrices $C_1$ and $C_2$ which satisfy the same properties as $C$ in the previous calculation and have dimensions $\beta_1$ and $\beta_2$ respectively. Suppose only the $n_k$ modes below the cutoff are involved. Further, recall that these dimensions are defined as the exponents of the non-interacting theory, so the four-point function can be broken into two-point functions.
\begin{equation*}
    \sum_{a,b<n_k} (C_1 \circ C_2)_{aabb}\langle \phi_a^2 \phi_b^2\rangle \sim \left(\int^k_0 d\lambda\, \rho(\lambda) \lambda^{1-\beta_1} \right)\left(\int_0^k d\lambda\, \rho(\lambda) \lambda^{1-\beta_2}\right) + \text{sub-leading terms} \sim k^{2\alpha + \beta_1 +\beta_2 - 2}
\end{equation*}
And so the engineering dimensions of couplings like $C_1 \circ C_2$ are just the sum of the dimensions of $C_1$ and $C_2$, that is:
\begin{equation*}
    [C_1 \circ C_2] = \beta_1 + \beta_2\,.
\end{equation*}
The dimensions of couplings in the energy can be found using these rules. As an example, consider the energy of the finite $\phi^4$ model, analyzed in the first section:
\begin{equation*}
    \mathcal{H}(\phi) = \frac{1}{2} \sum_{ab}(K_{ab} + u_2 \delta_{ab}) \phi_a\phi_b + \frac{u_4}{4!}\phi_a^2 \phi_b^2
\end{equation*}
The dimension of $\mathcal{H}$ must be $\alpha$:
\begin{equation*}
    [\mathcal{H}] = [K\cdot\phi^2] = 1 + 2[\phi] = \alpha
\end{equation*}
Following the rules above, we further find:
\begin{equation*}
    [u_2] = 1, \quad [u_4] = 2-\alpha
\end{equation*}

We reiterate that this method is heuristic, and rescaling by these dimensions will not always give scale-free flows. The primary reasons why this calculation may fail are anomalous scaling, the presence of operators like $\tilde{K}^z$ (which will be defined in the next section) which do not commute with the propagator $G$. When this latter situation is the case, $\tilde{K}^z$ is not diagonal in the collective variable basis, and so the $k$-scaling dimensions of loop sums like $\Tr [\dot{R}_k G_k \tilde{K}^A G_k \tilde{K}^A G_k]$ are not simply given by the sum of dimensions of operators under the trace. We are fortunate in this work that although this non-commutativity is present, the $\tilde{K}^z$ matrices appear to be approximately or effectively diagonal in the collective basis, and the loop sums have scaling dimensions which can be na\"ively approximated.

A third mechanism which can break this heuristic is the presence of operators which are not scale-free. One example could be if $K$ were not scale free, causing the eigenvalue density $\rho(\lambda)$ to not have simple power-law dependence on $\lambda$. This can cause the scaling dimensions to change along the flow \cite{lahoche_generalized_2021} and demonstrates why it is necessary to define the true scaling dimensions in terms of the flow equations and not \textit{a priori}. This contingency also demonstrates the power of NPRG formalism, since the flow equations are defined whether or not a scale-free description is available.

\subsubsection{RG calculation}

In the coalesced phase, the separation $s$ between $s_A$ and $s_B$ is smaller than a critical value $s_c$, which is calculated in the next section. Here we set $|s| = 0$ which significantly simplifies the analysis. Expanding \eqref{eq:mixture_model} with $s_A = s_B$ about the point $\phi=0$ yields at the quadratic level,
\begin{align*}
    \mathcal{H}^{(2)} &= \frac{1}{2}\left( K^A + K^B + (u^A_2 + u^B_2)\delta\right)\\
    &= K + u_2\delta\,.
\end{align*}
In the above expressions, we use $\delta$ to denote the identity matrix. Let us first consider the quadratic part, which we use to define scale. The density of eigenvalues of $K$ is given by
\begin{equation*}
    \rho(\lambda) = \sum_{a=1}^N\delta(\lambda - \lambda_a)\,,
\end{equation*}
which for large $N$ can be treated as an effectively smooth distribution, with proper care. Approximating these eigenvalues to be power-law distributed, define the scaling dimension $\alpha$ as
\begin{equation*}
    \alpha(k) = \partial_{\log k} \log \int_0^k d\lambda\, \rho(\lambda) \approx \alpha.
\end{equation*}
Next, We can calculate the scaling dimensions of $K^A$ and $K^B$ as discussed in the previous section. When written in the $K$-eigenbasis, $K^A$ and $K^B$ appear as approximately diagonal. The precise reasons for this are beyond the scope of this work, so we merely take it as an experimental fact. We must be careful to note that these matrices are not truly diagonal in the $K$-eigenbasis. Additionally, $K^A$ and $K^B$ display mass gaps in their diagonal elements just as $K$ does. We define their scale-free parts by subtracting off these mass gaps:
\begin{align*}
    \tilde{K}^A &= K^A - K^A_0 \delta \\
    \tilde{K}^B &= K^B - K^B_0 \delta
\end{align*}
This allows us to compute the $k$-scaling dimension of $\tilde{K}^A$, which we also approximate as constant with respect to $k$
\begin{equation*}
    [\tilde{K}^A] = \beta_A(k) = \partial_{\log k}\log\int_0^k d\lambda\, \sum_{a}\delta\left(\lambda - \diag_a\{\tilde{K}^A\}\right) \approx \beta_A,
\end{equation*}
and similarly for $\tilde{K}^B$, where by $\diag \{\tilde{K}^A\}$ we mean the diagonal element in the $K$-eigenbasis corresponding to the $\lambda + K_0$ eigenvalue of $K$.

In the topmost plot of Fig.\ \ref{fig:scaling_dimensions}, we show that by restricting to $\alpha_A = \alpha_B$ and sweeping, $\beta_A$ and $\beta_A$ are essentially unity, and that $\alpha$ has a predictable functional dependence. While small deviations from power-law behavior may be present, we approximate $\beta_A = \beta_B = 1$ for the rest of the analysis, unless otherwise specified.

\begin{figure}
    \centering
    \includegraphics[width=0.6\textwidth]{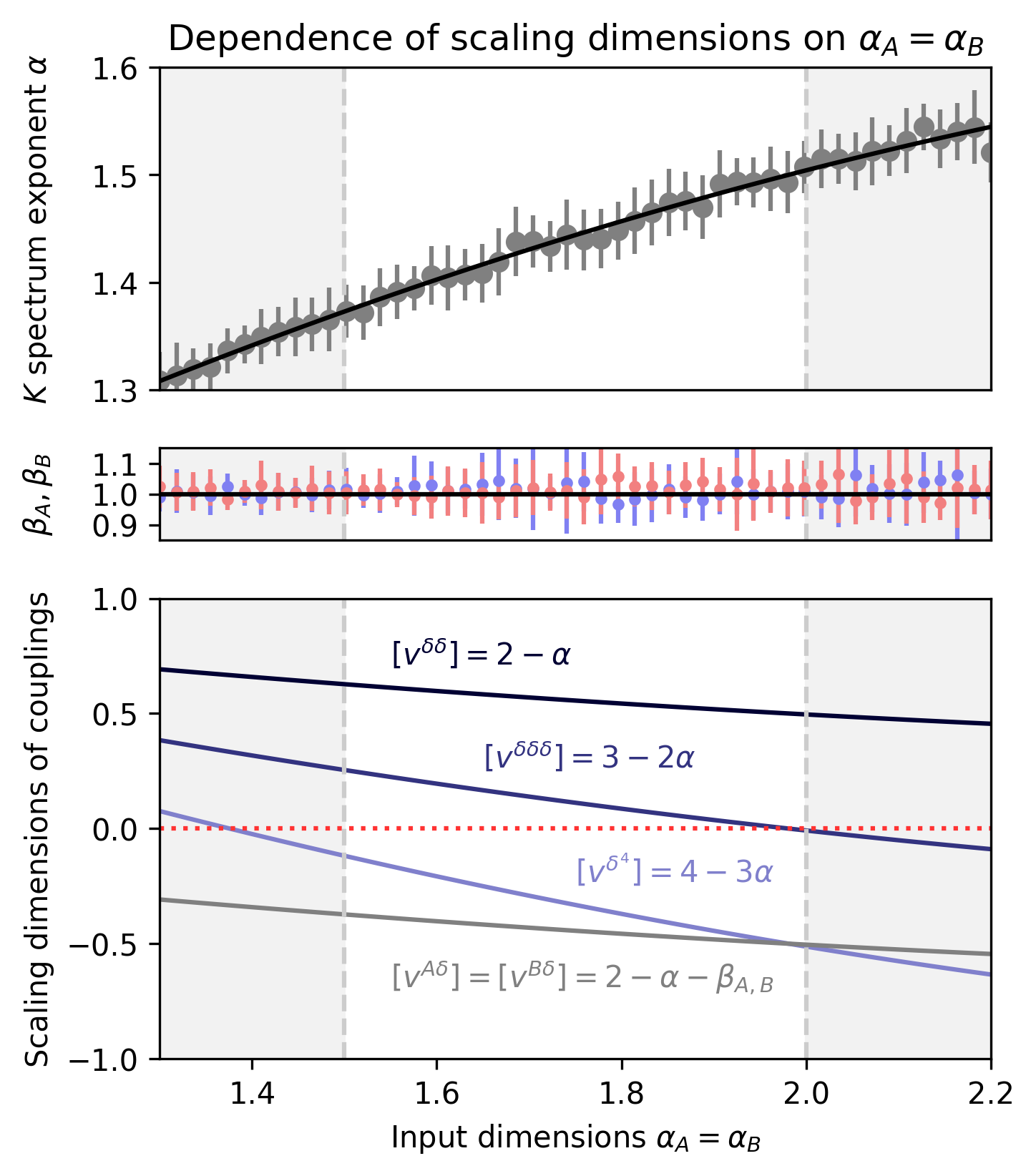}
    \caption{Numerical examination of the exponent $\alpha$ describing the scaling properties of eigenvalues of $K$. Areas without grey shading denote  $\alpha_A=\alpha_B\in(1.5,2)$, which corresponds to $D \in (3,4)$. \textbf{Top}: Estimates of the exponent $\alpha$ were obtained by explicitly constructing $K$ and measuring power laws in the scale-free part of their spectra. Error bars represent the root mean square error over 1\,000 trials at each value of $\alpha_A$. Matrix size is $N=75$. Black line is a polynomial fit which we use to find scaling dimensions. \textbf{Middle}: Estimates of $\beta_A$ and $\beta_B$ show no significant deviation from unity. \textbf{Bottom}: scaling dimensions of various couplings. The only couplings which are relevant in the range $\alpha_A=\alpha_B\in(1.5,2)$ are $v^{\delta\delta}$ and $v^{\delta\delta\delta}$}
    \label{fig:scaling_dimensions}
\end{figure}

We now turn our attention to the quartic couplings, which will contain most of the rest of the building blocks of our ansatz.
\begin{align*}
    \mathcal{H}^{(4)} &= -\frac{3}{4} K^A \circ K^A - \frac{3}{4} K^B \circ K^B + \frac{3}{2} K^A \circ K^B \\
    &- \frac{3 (u_2^A - u_2^B)}{2} K^A \circ \delta - \frac{3(u_2^B - u_2^A)}{2} K^B \circ \delta\\
    &+ \left(\frac{u_4^A}{2} + \frac{u_4^B}{2} - \frac{3((u_2^A)^2 + (u_2^B)^2)}{4}\right)\delta \circ \delta\,.
\end{align*}
The use of the symbol $\circ$ here denotes a symmetrized outer product:
\begin{align*}
    (K^A\circ K^A)_{abcd} &= K^A_{(ab}K^A_{cd)} \\
    &= \frac{1}{4!}\sum_{\text{perms}\, p}K^A_{p_1 p_2}K^{A}_{p_3 p_4}\,.
\end{align*}
Again, it is useful to separate out the couplings into scale-free parts. This yields:
\begin{align*}
    \mathcal{H}^{(4)} &= -\frac{3}{4} \tilde{K}^A \circ \tilde{K}^A - \frac{3}{4} \tilde{K}^B \circ \tilde{K}^B + \frac{3}{2} \tilde{K}^A \circ \tilde{K}^B \\
    &+ \frac{3}{2} (u_2^B + K^B_0 - u_2^A - K^A_0) \tilde{K}^A \circ \delta \\
    &+ \frac{3}{2} (u_2^A + K^A_0 - u_2^B - K^B_0) \tilde{K}^B \circ \delta\\
    &+ \left(\frac{u_4^A}{2} + \frac{u_4^B}{2} - \frac{3}{4}\left(u_2^A + K_0^A - u_2^B - K_0^B\right)^2\right)\delta \circ \delta\,.
\end{align*}

Finally, for the sixth-order part, we discard all but the $\delta\circ\delta\circ\delta$ term, which we shall justify shortly.
\begin{align*}
    \mathcal{H}^{(6)}_{a\dots f} &= \frac{15}{4} (u_4^A-u_4^B) \times \\ &\qquad(u_2^B-K_0^B-u_2^A+K_0^A)(\delta\circ\delta\circ\delta)_{a\dots f}
\end{align*}

For the moment, we explicitly include all of these operators in our ansatz for the effective action $\Gamma_k$. This gives at least ten couplings in total; three quadratic operators, six quartic, and one sixtic:
\begin{align}
    \Gamma_{k\,ab}^{(2)} &= (v^A_k \tilde{K}^A + v^B_k \tilde{K}^B + v^\delta_k \delta)_{ab} \label{eq:quadratic_ansatz_hard} \\
    \Gamma_{k\,abcd}^{(4)} &= (v^{AA}_k \tilde{K}^A\circ \tilde{K}^A + v^{AB}_k \tilde{K}^A\circ \tilde{K}^B + \nonumber\\
        &\qquad  v^{BB}_k \tilde{K}^B\circ \tilde{K}^B + v^{A\delta}_k \tilde{K}^A\circ \delta\, + \nonumber\\
        &\qquad v^{B\delta}_k \tilde{K}^B\circ \delta\, + v^{\delta\delta}_k \delta\circ\delta)_{abcd} \label{eq:quartic_ansatz_hard}\\
    \Gamma^{(6)}_{k\,a\dots f} &= (\dots + v^{\delta \delta \delta}_k \delta \circ \delta \circ \delta)_{a\dots f} \label{eq:sixtic_ansatz_hard}
\end{align}

Many of these couplings are irrelevant for $\alpha_A = \alpha_B \in (1.5,2)$, including all sixtic couplings except $v^{\delta \delta \delta}$. This follows from the dimensional analysis rules we discussed, together with our definition of $\alpha$.
\begin{align*}
    [\phi] &= \frac{1}{2}(\alpha - 1) ,\quad [\mathcal{H}^{(n)}] = \alpha\left(1-\frac{n}{2}\right) + \frac{n}{2}\\
    [v^A] &= 1 - \beta_A = 0, \quad [v^B] = 1 - \beta_B = 0, \quad [v^\delta] = 1\\
    [v^{z_1z_2}] &= 2 - \alpha - \beta_{z_1} - \beta_{z_2} \,;\quad z_i\in \{A,B,\delta\},\quad \beta_\delta = 0\\
    [v^{\delta \delta \delta}] &= 3 - 2\alpha
\end{align*}

A few of these scaling dimensions are represented as a function of $\alpha_A = \alpha_B$ in Fig. \ref{fig:scaling_dimensions}. Terms like $\tilde{K}^A\circ\delta$ and $\tilde{K}^A\circ\tilde{K}^B$ have negative dimensions for the range of spectral exponents we are considering, so these are termed irrelevant. Such terms can be important for obtaining accurate estimates of non-universal properties in terms of the original couplings, but deep in the IR, they only have the effect of shifting values of relevant couplings. As discussed, these dimensions are derived heuristically, and must be verified.

The ansatz (\ref{eq:quadratic_ansatz_hard}-\ref{eq:sixtic_ansatz_hard}) generates an approximately closed set of flow equations given the truncation, since the new couplings generated in the flow are driven only by sub-extensive contributions. As an example, we verify the predicted value of $s_{z_1 z_2} = [v^{z_1 z_2}_k]$ above. Looking only at the diagram with two 4-point vertices and keeping only extensive terms (see first section for explanation),
\begin{equation*}
    \partial_k v_k^{z_1 z_2} = \frac{1}{3} \sum_{\substack{{x \in \{A,B,\delta\}}\\y \in \{A,B,\delta\}}}v^{z_1 x}_k v^{z_2 y}_k\Tr\left[\dot{R}_k G_k C_{x} G_k C_{y} G_k\right]\,;\quad C_A = \tilde{K}^A,\, C_B = \tilde{K}^B,\, C_\delta = \delta
\end{equation*}
The LHS must have dimension $s_{z_1z_2} - 1$, while the RHS is a little trickier. Consider a single term:
\begin{equation}
    s_{z_1 z_2} = 1 + s_{z_1 x} + s_{z_2 y} + [\Phi_k]\,; \quad \Phi_k^{xy} = \Tr \left[\dot{R}_k G_k C_{x} G_k C_{y} G_k\right] \label{eq:quartic_renorm_scaling}
\end{equation}
In reality, these $C$ matrices do not generally commute with $G_k$. Yet, given the setup of our problem, the $\tilde{K}^z$ matrices effectively act as if they are diagonal when in the $K$ eigenbasis. The catch is that their diagonals are not eigenvalues and the scale free parts are no longer described by the exponents $\alpha_A$ and $\alpha_B$, but instead with $\alpha(\alpha_A,\alpha_B)$, as can be easily numerically verified. Therefore,
\begin{equation*}
    [\Phi^{xy}_k] = -3 + \alpha + \beta_x + \beta_y
\end{equation*}
Which shows, after direct substitution into \eqref{eq:quartic_renorm_scaling} that the free-field guess at $s_{z_1z_2}$ is correct.

The ansatz (\ref{eq:quadratic_ansatz_hard}-\ref{eq:sixtic_ansatz_hard}) can be used to find flow equations for all couplings, and these can be integrated numerically. At present we are only interested in asymptotic properties, so we can thin out the effective action ansatz by removing irrelevant terms. This takes us from ten down to five couplings:
\begin{align}
    \Gamma_{k\,ab}^{(2)} &= (v^A_k \tilde{K}^A + v^B_k \tilde{K}^B + v^\delta_k \delta)_{ab} \label{eq:quadratic_ansatz}\\
    \Gamma_{k\,abcd}^{(4)} &=  v^{\delta\delta}_k (\delta\circ\delta)_{abcd} \label{eq:quartic_ansatz}\\
    \Gamma^{(6)}_{k\,a\dots f} &= v^{\delta \delta \delta}_k (\delta \circ \delta \circ \delta)_{a\dots f} \label{eq:sixtic_ansatz}
\end{align}

Because $v^{A\delta}$ and $v^{B\delta}$ are irrelevant and have been dropped, $v^A$ and $v^B$ are not significantly renormalized by any couplings in this ansatz. The dimensionless flow equations for relevant parameters, computed using the Litim regulator and not accounting for anomalous scaling, are given by:
\begin{align*}
    \dot{g}^\delta_l &= g^{\delta}_l + \frac{1}{6} \frac{g^{\delta\delta}_l}{(1+g^{\delta}_l)^2}\\
    \dot{g}^{\delta\delta}_l &= (2-\alpha)g^{\delta\delta}_l - \frac{1}{3} \frac{(g^{\delta\delta}_l)^2}{(1+g^{\delta}_l)^3} + \frac{1}{10}\frac{g^{\delta\delta\delta}}{(1 + g^\delta)^2}\\
    \dot{g}^{\delta\delta\delta}_l &= (3-2\alpha)g^{\delta\delta\delta}_l + \frac{5}{3} \frac{(g_l^{\delta\delta})^3}{(1+g_l^\delta)^4} - \frac{g_l^{\delta\delta}\,g_l^{\delta\delta\delta}}{(1+g_l^\delta)^3}
\end{align*}

\subsection{\label{sec:multirelevance_breakdown_appendix}Calculation of multi-relevance breakdown separation $s_c$}
\subsubsection{Mean-field saddle point method}
To begin, assume the $A$ and $B$ models have all the same parameters, e.g. $u_2^A = u_2^B$, etc., but the eigenbases of $K^A$ and $K^B$ are random with respect to each other. (Each drawn from the Gaussian Orthogonal Ensemble.) The saddle point state is given by:
\begin{equation*}
    \phi_{\text{sp}} = \frac{1}{2}M^{-1}\,(M^B-M^A)\cdot s + O(|s|^3)
\end{equation*}
Where $M^A = K^A + u_2^A$ and $M = M^A + M^B$. The truncation in powers of $|s|$ is due to the fact that $s_c$ should go as $\Lambda^{-1/2}$, which we take to be small.  Set $s_A = |s|/2$ and $s_B=-s_A$ without loss of generality. The saddle point condition is approximately satisfied at the origin:
\begin{equation*}
    \frac{1}{2}M^{-1}\,(M^B - M^A)\cdot s \approx \frac{1}{2}M^{-1}
    \left(\frac{1}{N}\Tr M^B - \frac{1}{N}\Tr M^A\right) |s| \approx 0
\end{equation*}

An important approximation we made here is that since $s$ is randomly oriented with respect to the eigenbases of $K^A$ and $K^B$,  it is approximately an eigenvector for the ranges $\alpha$ we are dealing with, as can be verified numerically. Explicitly,
\begin{equation*}
    M^A \cdot s \approx \frac{1}{N} \left(\Tr M^A \right)|s|\,.
\end{equation*}

At the origin, the energy Hessian up to $O(|s|^4)$ is
\begin{align}\label{eq:breakdown_hessian}
    \mathcal{H}^{(2)}_{ab} &= M_{ab} - \frac{1}{4} (M\cdot s)_a(M\cdot s)_b + \frac{1}{2^4 3}(u_4^A + u_4^B) ( s^2 \delta_{ab} + 2 s_A s_B) + O(|s|^4) \\
    &= M_{ab} + \frac{1}{2^3 3} u_4  s^2 \delta_{ab} - \frac{1}{4}\left(\left(\frac{1}{N}\Tr M \right)^2 - \frac{1}{3}u_4\right) s_As_B
\end{align}
Though $ s$ is not an exact eigenvector of this Hessian, we approximate it as one. Its `eigenvalue' can be found from the above:
\begin{equation*}
    (\mathcal{H}^{(2)} \cdot  s)_a = \left( \frac{1}{N}\Tr{M} + \frac{1}{2^3} u_4  s^2 - \frac{1}{2^2}\left( \frac{1}{N}\Tr M \right)^2  s^2 \right)s_A
\end{equation*}
Setting this to zero, we can solve for the point at which curvature along the separation vector between the basins goes from positive to negative:
\begin{align} \label{eq:phi_c_mft}
    s_c^2 = 4 \left(\left(\frac{1}{N}\Tr M\right)^2 - \frac{1}{2}u_4\right)^{-1}
\end{align}

\subsubsection{Mixture component variance calculation}

The alternative method for calculating the critical separation is to leverage the fact that our model is a mixture of densities:
\begin{equation*}
    P_z(\phi) \propto \exp(-\mathcal{H}_z(\phi))\,,\quad P(\phi) \propto \exp(-\mathcal{H}_A(\phi)) + \exp(-\mathcal{H}_B( s))
\end{equation*}
Because we have assumed all of the same parameters for the $A$ and $B$ states (except basis),
\begin{equation*}
    P(\phi) = \frac{1}{2} P_A(\phi) + \frac{1}{2} P_B(\phi)
\end{equation*}
To calculate the probability density of $\phi$ along $\hat{ s}$, the unit vector along the basin separation, we evaluate
\begin{align*}
    P(\hat{ s}\cdot \phi = s) &= \int d\phi\, \delta(s - \hat{ s} \cdot \phi) P(\phi)\\
    &= \frac{1}{2}P_A(\hat{ s}\cdot \phi = s) + \frac{1}{2}P_B(\hat{ s}\cdot \phi = s)
\end{align*}

In the independent phase, $P_A(\hat{ s}\cdot \phi)$ does not significantly overlap $P_B(\hat{ s}\cdot\phi)$. If $A$ and $B$ are in their disordered phases, these distributions will be Gaussian, while in their ordered phases they will each look like a mixture two Gaussians placed at $\pm \hat{ s}_0 |\langle\phi_0\rangle_z|$, where by $|\langle\phi_0\rangle_z|$ we simply mean the magnetization in the standard sense. In both cases the coalescence transition occurs roughly when the distance between the centers of these distributions is twice the standard deviation of one of them, causing overlap:
\begin{equation}\label{eq:phi_c_rg}
    \left(\frac{s_c}{2}\right)^2 = \langle (\hat{ s} \cdot \phi)^2\rangle_{z} = \sum_{ab}\hat{ s}_a \hat{ s}_b \left(\Gamma_{z,k\to 0}^{(2)}\right)^{-1}_{ab}
\end{equation}
From the ansatz for $\Gamma_k$,
\begin{equation*}
    \left(\Gamma_{z,0}^{(2)}\right)_{ab} = (\lambda^z_a + u_{2,0}^{z})\delta_{ab}
\end{equation*}
In the $K^z$ eigenbasis. Using the fact that $ s$ has a random orientation in this basis,
\begin{equation*}
    \sum_{ab}\hat{ s}_a \hat{ s}_b \left(\Gamma_{z, 0}^{(2)}\right)^{-1}_{ab} \approx \frac{1}{N} \Tr \left(\Gamma_{z, 0}^{(2)}\right)^{-1}
\end{equation*}
In the disordered phase, $u^z_{2,0} > 0$, so
\begin{equation*}
\frac{1}{N} \Tr \left(\Gamma_{z, 0}^{(2)}\right)^{-1} = \frac{1}{N}\int_0^{\Lambda_z} d\lambda\, \rho_z(\lambda) \frac{1}{\lambda + u_{2,0}^z}
\end{equation*}
In the disordered phase, $u_{2,0}^z < 0$, so expand the potential instead around one of the minima at $\langle\phi_0\rangle = \pm (-6u_{2,0}^z/u_{4,0}^z)^{1/2}$. The total variance is $\langle\phi_0\rangle^2$ plus trace the inverse of $\Gamma^{(2)}$ evaluated at the symmetry-broken expansion point:
\begin{equation*}
    \sum_{ab}\hat{ s}_a \hat{ s}_b \left(\Gamma_{z, 0}^{(2)}\right)^{-1}_{ab} \approx \frac{1}{N}\left( -\frac{6 u_{2,0}^z}{u_{4,0}^z} + \int_0^{\Lambda_z} d\lambda\, \rho_z(\lambda) \frac{1}{\lambda - 2 u_{2,0}^z}\right)
\end{equation*}
There is a normalization factor on $\rho_z$ so that it integrates to $N$. For large $N$, our assumption is that
\begin{equation*}
    \rho_z(\lambda) \to \frac{N \alpha_z}{\Lambda_z^{\alpha_z}} \lambda^{\alpha_z-1}
\end{equation*}
Hence the disordered phase gives
\begin{equation*}
    \left(\frac{s_c}{2}\right)^2 \approx \frac{\alpha_z}{\Lambda_z^{\alpha_z}} \int_0^{\Lambda_z} d\lambda\,\frac{\lambda^{\alpha_z-1}}{\lambda + u_{2,0}^{z}}
\end{equation*}
While the ordered phase gives
\begin{equation*}
\left(\frac{s_c}{2}\right)^2 \approx -\frac{6}{N}\frac{u_{2,0}^z}{u_{4,0}^z}  + \frac{\alpha_z}{\Lambda_z^{\alpha_z}} \int_0^{\Lambda_z} d\lambda\,\frac{\lambda^{\alpha_z-1}}{\lambda + u_{2,0}^{z}}
\end{equation*}
\end{widetext}

\end{document}